\documentclass[aps,pra,preprint, superscriptaddress,showpacs]{revtex4-1}
\usepackage{bm}
\usepackage{color}
\usepackage[dvips]{graphicx}

\usepackage{graphicx}%

\usepackage{multirow}
\usepackage{setspace}
\usepackage{array}
\usepackage{amsmath}
\usepackage{booktabs}
\usepackage{tabularx}
\usepackage{enumerate}
\usepackage{dcolumn}% Align table columns on decimal point

\usepackage{amsfonts}
\usepackage{float}
\usepackage{subfigure}
\def\jcp#1#2#3{J.~Chem.~Phys.~{\bf #1},\ #2\ (#3)}

\def\pra#1#2#3{Phys.~Rev.~A~{\bf #1},\ #2\ (#3)}
\def\prl#1#2#3{Phys.~Rev.~Lett.~{\bf #1},\ #2\ (#3)}

\def\rmp#1#2#3{Rev.~Mod.~Phys.~{\bf #1},\ #2\ (#3)}

\def\aamop#1#2#3{Adv.~At.~Mol.~Opt.~Phys.~{\bf#1},\ #2\ (#3)}

\def\LA{L_{A}}
\def\LB{L_{B}}
\def\SA{S_{A}}
\def\SB{S_{B}}
\def\jA{j_{A}}
\def\jB{j_{B}}

\def\k1{k_1}
\def\k2{k_2}
\def\q1{q_1}
\def\q2{q_2}

\def\({\left (}
\def\){\right )}
\def\[{\left [}
\def\]{\right ]}

\def\SA{S_{A}}
\def\NA{N_{A}}

\newcommand{\beq}{\begin{equation}}
\newcommand{\eeq}{\end{equation}}

\newcommand{\threejm}[6]{ \left(\begin{array}{ccc} #1 & #3 & #5\\
                                             #2 & #4 & #6
                               \end{array}
                         \right)}
\newcommand{\sixj}[6]{ \left\{\begin{array}{ccc} #1 & #3 & #5\\
                                             #2 & #4 & #6
                               \end{array}
                         \right\}}

\begin{document}
\date{\today}
\title{
Efficient method for quantum calculations of molecule - molecule scattering properties in a magnetic field}  
% and prospects for evaporative cooling
\author{Y. V. Suleimanov}
\affiliation{Department of Chemical Engineering, Massachusetts Institute of Technology, Cambridge, Massachusetts 02139}%\email[]{ysuleyma@mit.edu}
\affiliation{Department of Mechanical and Aerospace Engineering, Combustion Energy Frontier Research Center, Princeton University, Princeton, New Jersey 08544}\email[]{yury.suleymanov@gmail.com}
\author{T. V. Tscherbul}
\affiliation{Harvard-MIT Center for Ultracold Atoms, Cambridge, Massachusetts 02138}
\affiliation{Institute for Theoretical Atomic, Molecular, and Optical Physics,
Harvard-Smithsonian Center for Astrophysics, Cambridge, Massachusetts 02138}%\email[]{tshcherb@cfa.harvard.edu}
\author{R. V. Krems}
\affiliation{Department of Chemistry, University of British Columbia, Vancouver, BC V6T 1Z1, Canada}
%\email[]{rkrems@chem.ubc.ca}

\begin{abstract}
We show that the cross sections for molecule - molecule collisions in the presence of an external field can be computed efficiently using a total angular momentum basis, defined either in the body-fixed frame or in the space-fixed coordinate system. This method allows for computations with much larger basis sets than previously possible. We present calculations for $^{15}$NH - $^{15}$NH collisions in a magnetic field. Our results support the conclusion of the previous study that the evaporative cooling of rotationally ground $^{15}$NH molecules in a magnetic trap has a prospect of success.

\end{abstract}

\maketitle

\clearpage
\newpage

\section{Introduction}

The creation of cold and ultracold molecular ensembles has 
opened new opportunities for fundamental research in physics and physical chemistry  \cite{NJP,Book:PrecisionTests,Book:Dipoles,Softley,SchnellMeijer,Dulieu,JohnCPC}. 
The rich internal structure of molecules represented by nested manifolds of electronic, vibrational, rotational, and hyperfine levels \cite{Herzberg} can be used to engineer intermolecular interactions at low temperatures \cite{Roman08}, enabling new schemes for quantum information processing \cite{QIP}, quantum simulation of spin-lattice Hamiltonians \cite{QuantumSimulation,felipe2011,Zoller1,Zoller2}, precision measurements of fundamental physical constants \cite{Book:PrecisionTests}, external field control of bimolecular collision dynamics  \cite{KRb2,OH-ND3,N-NH} and high-resolution measurements of collision-induced energy transfer probabilities \cite{RgOH}. Ultracold molecules can be produced by photoassociation of laser-cooled alkali-metal atoms \cite{Dulieu} leading to the formation of ultracold alkali-metal dimers \cite{KRb}. A variety of alternative experimental techniques including buffer-gas cooling \cite{BufferGasCooling}, laser cooling \cite{LaserCooling},  velocity filtering \cite{Rempe} and Stark and Zeeman deceleration of molecular beams \cite{StarkDeceleration,ZeemanDeceleration} produce molecular ensembles at temperatures 10 - 500 mK. In order to reduce the temperature of the molecular gas below 10 mK, these techniques need to be supplemented with an additional (second-stage) cooling method \cite{NJP} such as 
sympathetic cooling with ultracold atoms \cite{NJP,He-CH2,Lara,Gianturco,Wallis,Barletta} or evaporative cooling \cite{NJP,Ketterle}. 

The evaporative cooling experiments rely on elastic collisions, which return the gas to thermal equilibrium after the most energetic particles are removed from the ensemble \cite{Ketterle}.  If evaporative cooling is performed in a magnetic trap, collisions between molecules in low-field-seeking Zeeman states can lead to electron spin depolarization, causing trap loss \cite{Roman03,Roman05,prl09,N-NH-PCCP}.
 An empirical rule indicates that evaporative cooling works if elastic collisions occur $\gamma \sim 100$ times more frequently than inelastic collisions \cite{NJP}. 
While evaporative cooling is widely used to create ultracold atomic gases \cite{Ketterle}, 
the applicability of this technique to molecules in low-field-seeking states remains an open question. Molecular collisions at low temperatures are determined by strongly anisotropic interactions, which may lead to large inelastic collision rates. In order to predict whether or not evaporative cooling of specific molecules would be possible, it is necessary to determine the cross sections for elastic and inelastic scattering in molecule - molecule collisions in the presence of the trapping fields. Unfortunately, the fully quantum calculation of molecule - molecule scattering properties in external fields remains a daunting task. Selected calculations have been reported for OH($^2\Pi$) - OH($^2\Pi$) collisions in an electric field \cite{AvdeenkovBohn}, O$_2(^3\Sigma_g^-)$ - O$_2(^3\Sigma_g^-)$ collisions in a magnetic field \cite{NJP09} and NH($^3\Sigma$) - NH($^3\Sigma$) collisions in a magnetic field \cite{JanssenPRA11}. These calculations are based on the fully uncoupled space-fixed (SF) representation of the scattering wave functions \cite{VolpiBohn,Roman04}, which results in prohibitively large basis sets and precludes fully converged computations with realistic intermolecular potentials. The basis set truncation error of such computations is often difficult to estimate. 

Recently, Tscherbul and Dalgarno \cite{jcp10} showed that quantum calculations of atom - molecule scattering properties in an external field can be performed using an expansion of the scattering wavefunction in eigenfunctions of the total angular momentum $J$ of the collision complex in the body-fixed (BF) coordinate frame \cite{jcp10}. While $J$ is not conserved in the presence of an external field, the total angular momentum representation allows for a much more efficient truncation of the basis set than the uncoupled representation \cite{Li-CaH}. In the present work, we use the total angular momentum representation to extend the $^{15}$NH - $^{15}$NH scattering calculations of Janssen et al \cite{JanssenPRA11} in order to compute the scattering cross sections with a much larger basis set including up to 7 rotational states of $^{15}$NH.   
The imidogen radical (NH)  is currently studied in several experiments aiming at the production of ultracold molecules \cite{N-NH,N-NH-PCCP,BufferGasCoolingNH,StarkDecelerationNH-PRA,NHlifetime,NNHcotrapping}. Our results demonstrate that the method developed in Ref. \cite{jcp10} can be applied to molecule - molecule scattering after a suitable modification to allow for the exchange symmetry of the two-molecule wave function. Our calculations corroborate the conclusions of Janssen et al \cite{JanssenPRA11} and remove the uncertainties due to the basis set truncation errors. Our results demonstrate the effect of increasing the basis set on molecule - molecule scattering in a magnetic field at low temperatures and can be used as benchmark for future calculations.

\section{Theory}

There are two equivalent formulations of the scattering theory that can be used for the calculations presented here. The body-fixed formulation introduced in Ref. \cite{jcp10} is based on representing the scattering wave function ($\psi$) of two molecules by an expansion in terms of BF basis functions. This basis set is convenient for evaluating the matrix elements of the intermolecular interaction potential, but leads to complicated expressions for the matrix elements of the angular momentum operator that describes the rotation of the collision complex in the laboratory frame and the molecule - field interaction operators. An alternative formulation is based on representing the scattering wave function $\psi$ by an expansion in terms of eigenfunctions of the total angular momentum operator defined in the space-fixed frame 
and evaluating the matrix elements of the Hamiltonian directly in the SF basis.  
The SF and BF basis sets are related by a unitary transformation.  The numerical efficacy of these two basis sets is the same and the results obtained with these two basis sets must be identical. 
We present below both formulations. Calculating the scattering observables with these two different basis sets is usually a good test of the accuracy of the computations.

\subsection{Body-fixed formulation}

The Hamiltonian for a non-reactive collision of two $^3\Sigma $ molecules $A$ and $B$ may be written in the following form \cite{jcp10}
\begin{eqnarray}
\hat{H} & = & -\frac{1}{2\mu R}\frac{\partial^2}{\partial R^2} R + \frac{\hat{l}^2}{2\mu R^2}
 +  \hat{H}^{(A)}_\text{as} + \hat{H}^{(B)}_\text{as}  \notag \\
& +&   \hat{V}(R, \theta_A,\theta_B, \phi) +  \hat{V}_\text{dd}(R, \hat{S}_A, \hat{S}_B)  , \label{molBF1}
\end{eqnarray}
where $\mu $ is the reduced mass of the collision complex, $R$ is the separation between the centers of mass of the molecules, $ \hat{l} = (\hat{J} - \hat{N}_A - \hat{S}_A - \hat{N}_B - \hat{S}_B)$ is the orbital angular momentum describing the rotational motion of the collision complex, 
 $\hat{J}$ is the total angular momentum,  $\hat{N}_i$ and $\hat{S}_i$ are the rotational angular momentum and electronic spin, respectively, of the molecule $i$, and the sets of angles ($\theta_i,\phi_i$) describe the orientation of the molecules with respect to the BF quantization axis $z$ directed along $\hat{R}$. The hat over the symbol is used to denote angular momentum operators and unit vectors.

The Hamiltonian operators describing the separated molecules can be written as follows
\begin{equation}\label{molBF2}
\hat{H}^{(i)}_{\text{as}} = B_e \hat{N_i}^2 + \gamma_{\rm SR} \hat{N_i}\cdot\hat{S_i} + \frac{2}{3}\lambda_\text{SS} \left(\frac{24\pi}{5}\right)^{1/2}\sum_{q=-2}^2 (-)^q Y_{2,-q}(\theta _i, \phi _i) [\hat{S}_i\otimes\hat{S}_i]^{(2)}_q + 2\mu_0 {B} \hat{S}^{(i)}_Z,
\end{equation}
where $B_e$ is the rotational constant,  $\hat{S}^{(i)}_Z$ is the $Z$-component of $\hat{S}_i$, $\mu_0$ is the Bohr magneton, $\gamma_{\rm SR}$ is the spin-rotation interaction constant \cite{Mizushima}, $\lambda_\text{SS}$ is the spin-spin interaction constant\cite{Mizushima} and $[\hat{S}\otimes\hat{S}]^{(2)}_q$ is a spherical tensor product of $\hat{S}$ with itself \cite{Zare}.
The SF quantization axis $Z$ is chosen to point along the direction of the magnetic field. 

The operator describing the magnetic dipole-dipole interaction between the molecules is given by \cite{Roman04}  
\begin{equation}\label{molBF3}
\hat{V}_\text{dd}(R, \hat{S}_A, \hat{S}_B) = - g_s^2 \mu_0^2 \left( \frac{24\pi}{5}\right)^{1/2} \frac{\alpha^2}{R^3} \sum_q (-)^q Y_{2,-q} (\hat{R}) [\hat{\SA} \otimes \hat{\SB}]^{(2)}_q,
\end{equation}
where $g_s$ is the electron $g$-factor, $\alpha $ is the fine-structure constant and $[\hat{S}_A \otimes \hat{S}_B]^{(2)}_q$ is the spherical tensor product of $\hat{S}_A$ and $\hat{S}_B$ \cite{Zare}. Note that $g_s^2 \mu_0^2 \approx 1.0023~\hbar^2 e^2 /m_e^2$ \cite{NHNHfieldfree}. 
This factor, approximately equal to one when expressed in atomic units, was omitted from the corresponding expressions in Ref. \cite{Roman04}.

The interaction of two  molecules with electronic spins $S_A=S_B=1$ gives rise to three potential energy surfaces labeled by $S=0,1$ and 2, where $\hat{S}=\hat{S}_A+\hat{S}_B$.
As in the previous studies of $^{15}$NH-$^{15}$NH collisions \cite{JanssenPRA11,NHNHfree, NHNHfieldfree}, we assume that all three spin states of the $^{15}$NH-$^{15}$NH complex
 are described by the nonreactive
quintet surface, i.e. $V(R, \theta_A,\theta_B, \phi) = V_{S=2}(R, \theta_A,\theta_B, \phi) $ \cite{JanssenJCP09}.
The intermolecular interaction potential in Eq.~(\ref{molBF1}) can then be expanded in angular basis functions \cite{jcp10}
\begin{eqnarray}
V_S (R,\theta_A,\theta_B,\phi) & =  & (4\pi)^{3/2} \sum_{\lambda_A,\, \lambda_B,\, \lambda} \biggl{(}\frac{2\lambda+1}{4\pi}\biggr{)}^{1/2} V^{S=2}_{\lambda_A\lambda_B\lambda}(R) 
\notag \\ 
& &\times \sum_m \threejm{\lambda_A}{m}{\lambda_B}{-m}{\lambda}{0} Y_{\lambda_A m}(\theta_A,\phi_A)Y_{\lambda_B,-m}(\theta_B,\phi_B),\label{molBF4}
\end{eqnarray}
where $Y_{\lambda_i m}(\theta_i,\phi_i)$ are the spherical harmonics and the  parentheses denote the 3$j$-symbols \cite{Zare}. 
This expansion is particularly useful for evaluating the matrix elements of the interaction potential in the BF basis:  
\begin{equation} \label{molBF5}
|\Psi \rangle = \frac{1}{R}\sum_{\alpha_A,\alpha_B} \sum_{J,\Omega} F^M_{\alpha_A\alpha_B J\Omega}(R)
 |N_A K_{N_A}\rangle |S_A\Sigma_A\rangle   |N_B K_{N_B}\rangle |S_B\Sigma_B\rangle|JM\Omega\rangle,
 \end{equation}
where
$\Omega $, $K_{N_i}$, and $\Sigma _{i}$ are the projections of $\hat{J}$, $\hat{N}_i$, and $\hat{S}_i$ on the BF quantization axis $z$, and $M$ is the projection of $\hat{J}$ on the SF quantization axis $Z$. The symbols $\alpha_{A}$ and $\alpha_{B}$ denote collectively the quantum numbers $\{N_A,K_{N_A},S_A,\Sigma _A\}$ and $\{N_B,K_{N_B},S_B,\Sigma _B\}$ specifying the states of the separated molecules $A$ and $B$, respectively. 

In Eq.~(\ref{molBF5}),
\begin{equation} \label{molBF5a}
|JM\Omega \rangle \equiv \(\frac{2J+1}{8\pi ^2}\)^{1/2}D_{M\Omega }^{J*}(\Omega _E),
 \end{equation}
where $D_{M\Omega }^{J}$ is a Wigner $D$-function, describing the rotation of the collision complex with angular momentum $J$ in the SF frame \cite{Zare}. 

Ref.~\cite{jcp10} presents the matrix elements of the interaction potential (\ref{molBF4}) and the $\hat{l}^2$ operator in the BF basis~(\ref{molBF5}). The matrix elements of the asymptotic Hamiltonians (\ref{molBF2}) can be obtained in the same basis by recoupling the angular momenta as was done for the atom - molecule collision system in Ref.~\cite{jcp10}. 
The expression on the right-hand side of Eq.~(\ref{molBF3}) is a contraction of spherical tensor operators of the same rank and therefore it is independent of the choice of the coordinate frame. Setting $\hat{R}=0$ and using the Wigner-Eckart theorem \cite{Zare}, we obtain the matrix elements of the dipole - dipole interaction operator in the BF basis (\ref{molBF5})
\begin{eqnarray}
& & \langle JM\Omega| \langle N_A K_{N_A} | \langle S_A \Sigma_{A} | \langle N_B K_{N_B} | \langle S_B \Sigma_{B} | 
V_\text{dd} | J'M'\Omega'\rangle | N_A' K_{N_A}' \rangle | S_A \Sigma_{A}' \rangle | N_B' K_{N_B}'\rangle | S_B \Sigma_{B}'\rangle
\notag \\ 
& & = \delta_{JJ'}\delta_{MM'}\delta_{\Omega\Omega'} \delta_{N_A N_A'}\delta_{K_{N_A} K_{N_A}' } \delta_{N_B N_B'}\delta_{K_{N_B} K_{N_B}' } \left( \frac{-\sqrt{30} g_s^2 \mu_0^2 \alpha^2}{R^3} \right) \sum_{q_A,q_B} \threejm{1}{q_A}{1}{q_B}{2}{0} 
(-)^{S_A+S_B-\Sigma_A-\Sigma_B} \notag \\ 
&& \times [(2S_A+1)S_A(S_A+1)]^{1/2} [(2S_B+1)S_B(S_B+1)]^{1/2} \threejm{S_A}{-\Sigma_A}{1}{q_A}{S_A}{\Sigma_A'} 
\threejm{S_B}{-\Sigma_B}{1}{q_B}{S_B}{\Sigma_B'}.  \label{molBF8}
\end{eqnarray}

For collisions of identical molecules, the basis set in Eq.~(\ref{molBF5}) should be modified to account for the effects of the permutation symmetry. 
This  has been done by a number of authors in the SF frame (see, e.g. Refs. \cite{NJP09,JanssenPRA11,Roman04,Millardpaper,Kouripaper}).
In the case of  the BF total angular momentum representation given by Eq.~(\ref{molBF5}), the symmetrization can be accomplished by performing the transformation to the SF frame \cite{Zare} and applying the symmetrization operator $1+\hat{P}_{AB}$ to the right-hand side of Eq.~(\ref{molBF5}) \cite{Millardpaper,Kouripaper}. As a result, we obtain the symmetrized orthonormal BF basis states for two identical molecules
\begin{eqnarray}
& & |\phi ^\eta_{\alpha_A\alpha_B JM\Omega}\rangle  =  \frac{1}{[2(1+\delta_{\Omega,0}\delta_{N_A,N_B}\delta_{S_A,S_B}\delta_{K_{N_A}+K_{N_B},0})]^{1/2}} [ |N_A K_{N_A}\rangle |S_A\Sigma_A\rangle   |N_B K_{N_B}\rangle |S_B\Sigma_B\rangle|JM\Omega\rangle   \notag \\ 
&  & + \eta (-)^{N_A+S_A+N_B+S_B-J} |N_A {-K_{N_A}}\rangle |S_A-\Sigma_A\rangle   |N_B -K_{N_B}\rangle |S_B-\Sigma_B\rangle |JM-\Omega \rangle ],  \label{molBF7}
\end{eqnarray}
where $\eta= +1 (-1)$ for identical bosons (fermions), $\Omega\ge 0$, and we assume that the states are well-ordered. 

Eqs.~(\ref{molBF8}) and (\ref{molBF7}) and the expressions for the matrix elements in Ref. \cite{jcp10} can be combined to obtain the elements of the Hamiltonian matrix describing the interaction of indistinguishable molecules.

\subsection{Space-fixed formulation}

The matrix elements of the operator $\hat{l}^2$  in the BF basis~(\ref{molBF5}) can be lengthy and cumbersome to program. An alternative formulation of the collision problem can be derived using the total angular momentum representation directly in the SF coordinate frame. To do this, we re-write the expansion~(\ref{molBF5}) as 
\begin{equation} \label{molSF1}
|\Psi \rangle = \frac{1}{R}\sum_{\alpha_A,\alpha_B} \sum_{J} F^M_{\alpha_A\alpha_B J}(R)
 |JMjl(\jA \jB)\rangle,
 \end{equation}
where $\hat{j}_{ A}$ is defined as the vector sum of $\hat{N}_{A}$ and $\hat{S}_{A}$, 
$\hat{j}_{B}$  as the vector sum of $\hat{N}_{B}$ and $\hat{S}_{B}$, 
$\hat{j}$ as the vector sum of $\hat{j}_{A}$ and $\hat{j}_{B}$, and 
$\hat{J}$ as the vector sum of $\hat{j}$ and $\hat{l}$. The basis set representation (\ref{molSF1}) is most suitable for molecule - molecule scattering problems in the absence of external fields \cite{NHNHfieldfree} but it can also be used for computations of cross sections for molecular collisions in external fields. The matrix elements of the operator $\hat{l}^2$ in the SF representation (\ref{molSF1}) are particularly simple:
\begin{equation}
 \langle JMjl(\jA \jB)| \hat{l}^2 | JMj'l'(\jA' \jB')\rangle =
  \delta_{\jA \jA'} \delta_{\jB \jB'} 
 \delta_{ll'} \delta_{jj'} l(l+1). 
\end{equation}

The basis states (\ref{molSF1}) with a given value of $J$ form $J$-manifolds that lead to $J$-blocks of the Hamiltonian matrix as described in Ref. \cite{jcp10}.
 The blocks with different $J$ values are coupled by the interaction of the molecules with the external magnetic field given by the last term in Eq. (\ref{molBF2}). To evaluate the matrix elements of this operator in the basis (\ref{molSF1}), we note that the operators $\hat{S}^{ (i)}_{Z}$ can be represented by spherical tensors of rank 1 ($\hat{S}^{ (i)}_{Z} = \hat{T}^1_{q=0}$) and use the Wigner-Eckart theorem \cite{Zare}
\begin{eqnarray}
 \langle JMjl(\jA \jB) | \hat{S}^{(i)}_{Z} | J' M j'l'(\jA' \jB') \rangle = 
 (-1)^{J-M} \threejm{J}{-M}{1}{0}{J'}{M}
 \langle J jl(\jA \jB) || \hat{S}^{(i)}_{Z} || J'  j'l'(\jA' \jB') \rangle .
\nonumber
\\
\end{eqnarray}

The operators $\hat{S}^{(i)}_{Z}$ act on the subspace of one molecule. The reduced matrix elements of the operator $\hat{S}^{(A)}_{Z}$ can be obtained using Eq. (5.72) of Ref. \cite{Zare} to yield
\begin{eqnarray}
 \langle J jl \jA (\NA \SA) \jB  || \hat{S}^{(A)}_{Z} || J'  j'l' \jA' (\NA' \SA') \jB' \rangle   =   \delta_{ll'} \delta_{\jB \jB'} \delta_{\NA \NA'}
(-1)^{J + l +j + j' + \jB' + \NA + \SA + 1} \times  & & \nonumber \\
\left [ (2J+1) (2J'+1) (2j'+1)(2j+1) (2 \jA + 1 ) (2 \jA' + 1 ) \right ]^{1/2}  & & \nonumber  \\
\sixj{j}{J'}{j'}{J}{1}{l}\sixj{\jA}{j'}{\jA'}{j}{1}{\jB}\sixj{\SA}{\jA'}{\SA}{\jA}{1}{\NA}, & & \nonumber \\
\end{eqnarray}
where the factors in the curly braces are $6j$ symbols. 

The matrix elements of the molecule - molecule interaction potential operator in the SF basis (\ref{molSF1}) can be obtained using the theory described in Ref. \cite{jpc2004}. In particular, if the molecule - molecule interaction potential for each spin multiplicity is represented by the expansion given in Eq. (3) of Ref. \cite{jpc2004}, then the matrix elements of the total interaction potential (including the Heisenberg exchange interaction \cite{NJP09}) in the total angular momentum basis (\ref{molSF1}) are given by Eq. (33) of Ref. \cite{jpc2004} after substituting $\LA$ and $\LB$ with $\NA$ and $N_{B}$. If the potentials of different spin multiplicity of the molecule - molecule complex are assumed to be identical, the matrix elements of the interaction potential can be evaluated using a simpler expression given in Eq. (A1) of Ref. \cite{NHNHfieldfree}.
Ref. \cite{NHNHfieldfree} also gives the expressions for the matrix elements of the operators (\ref{molBF3}) for zero magnetic field and the matrix elements of the magnetic dipole - dipole interaction (\ref{molBF3}) in a SF total angular momentum basis. 
 Note that the total angular momentum basis set used in Ref. \cite{NHNHfieldfree} is based on a different coupling scheme. However, additional recoupling transformation  required to bring the Hamiltonian matrix defined in Ref. \cite{NHNHfieldfree} to the basis set representation (\ref{molSF1}) can be readily obtained by a simple modification of Eq. (30) in Ref. \cite{jpc2004}. We recommend the use of the representation (\ref{molSF1}) over the total angular momentum basis used in Ref. \cite{NHNHfieldfree} because it allows one to construct the basis by concatenation of single-molecule basis sets. 
%To avoid the recoupling transformation, we invite the reader to re-derive the matrix elements of the interaction potential and the field-free molecular Hamiltonians directly in the basis representation (\ref{molSF1}). This can be done by expressing all operators in the Hamiltonian (\ref{molBF1}) as products of spherical tensors and applying the formalism described in Section (5.4) of Ref. \cite{Zare}. 
 
As mentioned in the previous section, the symmetrization procedure for collisions of identical molecules can be applied directly in the SF coordinate frame.  
Using the SF total angular momentum representation (\ref{molSF1}) as the basis, we can write the symmetrized orthonormal SF basis states for two identical molecules as follows \cite{Kouripaper}: 
\begin{eqnarray}
& &|\phi ^\eta_{\alpha_{A}\alpha_B JMjl}\rangle  = \frac{1}{[2(1+\delta _{\jA\jB})]^{1/2}}[|JMjl(\jA\jB)\rangle+\eta(-)^{\jA+\jB+j+l}|JMjl(\jB\jA)\rangle]. 
\end{eqnarray}

\section{Numerical Results}

For numerical computations, the basis sets (\ref{molBF1}) and (\ref{molSF1}) are constructed by first fixing the total angular momentum value $J$ and then including a certain number of rotational states $\NA$ and $N_{B}$ from $\NA = N_{B} = 0$ to $\NA = N_{B} = N_{\rm max}$. Given the values of $J$, $\NA$ and $N_{B}$, the complete set of other quantum numbers defining the basis sets (\ref{molBF1}) and (\ref{molSF1}) is generated. This gives a manifold of $J$ basis states. As described above, the interactions of molecules with the external field couple basis states with different values of $J$. The basis sets must therefore include several $J$-manifolds up to a chosen value of $J = J_{\rm max}$ simultaneously. 
The projection $M$ of $\hat{J}$ on the external field axis remains a good quantum number and the calculations are performed in a cycle over $M$.
Due to the magnetic field-induced couplings between different $J$-blocks, the truncation of the total angular momentum basis sets at finite $J = J_{\rm max}$ leads to 
the appearance of unphysical eigenstates \cite{jcp10}. The unphysical states arise from the block of the Hamiltonian matrix corresponding to the largest value of $J$ and do not affect the scattering calculations at low collision energies \cite{jcp10}. In this calculation, we simply disregard the unphysical states when we calculate the scattering $S$-matrix. 
The correct physical states can be readily detected as they have the same energies as the sum 
of the Zeeman levels of isolated molecules.

 The calculations presented in this work were obtained with a computer code based on the BF formulation of the scattering problem described above. 
The reader can obtain a copy of the Fortran code by contacting the authors. The accuracy of this code was verified by comparing the results of calculations with the small basis set ($N_{\rm max} = 2$) for two collision energies (10$^{-6}$ and 10$^{-5}$ cm$^{-1}$) and two magnetic field values (100 and 1000 Gauss) with the results obtained using the code based on the uncoupled SF representation developed in our previous study of the O$_2$-O$_2$ collisions \cite{NJP09}. The integration of the coupled differential equations was performed using the log-derivative method \cite{David} on a radial grid ranging from 4.5 $a_0$ to 500 $a_0$ in steps of 0.05 $a_0$. Convergence at low collision energies was controlled by evaluating the cross sections in steps of 25 $a_0$. 

In order to verify the accuracy of our method and computations, we also repeated the calculations of Janssen et al. \cite{JanssenPRA11} for $^{15}$NH - $^{15}$NH scattering at ultralow collision energies in a magnetic field of 100 Gauss using the total angular momentum representation. The calculations are for collisions of molecules initially in the lowest energy state of the manifold characterized by $m_{\jA} = m_{\jB} = + 1$. The Zeeman transitions to lower-energy states characterized by the projections $m_{\jA}$ and $m_{\jB}$ equal to $0$ and $-1$ lead to inelastic relaxation.  The basis set for this computation includes 5 $J$-blocks ($J_{\rm max} =4$) and a total of 3 rotational states for each molecule ($ N_{\rm max}$=2). The calculations performed by Janssen et al. using  the uncoupled SF representation \cite{Roman04}  included 3 rotational states and 6 partial waves (see Ref.~\cite{JanssenPRA11} for more details).  The results displayed in Figure 1 demonstrate that these independent calculations are in good agreement. The disagreement at higher collision energies is due to incomplete numerical convergence of both calculations. This underlines the importance of basis set convergence. When the basis set is large enough, the calculations using the coupled angular momentum basis must yield the same results as the calculations using the uncoupled basis. However, the two basis sets, by construction, are not completely equivalent (see Ref.~\cite{NHNHfieldfree} for details). Therefore, if the basis sets are severely restricted, the uncoupled basis representation may produce different results from those based on the total $J$ representation. 
The calculations presented in Figure 1 were performed with the following values of the constants parameterizing the Hamiltonian matrix: $B_e=16.27034 $ cm$^{-1}$, $\gamma_{\rm SR} = -0.05460 $ cm$^{-1}$, and $\lambda_\text{SS} = 0.91989$ cm$^{-1}$. All other calculations presented in this work were performed with $B_e=16.245 $ cm$^{-1}$, $\gamma_{\rm SR} = -0.05467 $ cm$^{-1}$, and $\lambda_\text{SS} = 0.9197 $ cm$^{-1}$. 

Janssen et al. used the uncoupled SF representation \cite{Roman04} introduced for collision problems in the presence of external fields. This restricted their computations to the basis set including only three rotational states for each molecule. The total angular momentum representations described in the previous section allow for scattering calculations with  a larger basis set than the uncoupled SF representation. In order to explore the effect of the basis set size on the molecule - molecule scattering cross sections at low energies, we present in Figure 2 the results obtained with different numbers of molecular rotational states in the basis set. The results show that both the elastic and inelastic scattering cross sections are extremely sensitive to the basis set size until $N_{\rm max}$ reaches $5$. Figure 3 displays the energy dependence of the cross sections computed with the different basis sets in the energy interval of interest for ultracold molecule experiments. 

The computation with the basis set constrained by $N_{\rm max} = 6$ and $J_{\rm max} = 4$ involves the numerical integration of $18852$ coupled differential equations. The same calculation would require the integration of 38170 equations in the uncoupled basis set used by previous authors (assuming that, as in their previous calculations, the number of partial waves is fixed to 6). Numerical integration of a system of 18852 coupled differential equations for low energy collision problems takes  approximately 2700 hours on the best available computer processor. The computation time increases cubically with the number of basis functions. 

The computations presented in Figures 2 and 3 were carried out with the fixed value of the total angular momentum projection $M=2$. In principle, the full quantum calculation requires a summation over the results computed with different values of $M$. However, the scattering of molecules at low collision energies is usually dominated by the contribution of the partial cross sections corresponding to a few (or even a single) values of $M$. Figure 4 demonstrates that the computations with the fixed value $M=2$ provide accurate results for the elastic and inelastic scattering of molecules in the $m_{\jA} = m_{\jB} = + 1$ state. 

  The possibility of evaporative cooling of molecules in a magnetic trap remains one of the most pressing questions in the research field of cold molecules. Magnetic fields confine molecules in low-field-seeking states such as the $m_{\jA} = + 1$ state of the ground rotational level of $^{15}$NH. 
It is generally believed that the anisotropy of the molecule - molecule interaction potentials must be very large, leading to prohibitively large rates of inelastic Zeeman relaxation in collisions of trapped $^3\Sigma $ molecules. The evaporative cooling is deemed possible when the ratio $\gamma$ of elastic scattering cross sections and cross sections for inelastic collisions exceeds 100. Figure 5 presents the magnetic field dependence of $\gamma$ for $^{15}$NH - $^{15}$NH collisions computed for several collision energies. In the limit of low collision energies, the elastic cross sections are energy independent and the inelastic cross sections are inversely proportional to the collision velocity \cite{wigner}. Therefore, $\gamma$ must decrease when the collision energy decreases. At the same time, the rate of spin relaxation tends to zero when the magnetic field lifting the degeneracy of the Zeeman levels vanishes \cite{VolpiBohn,threshold}. 
The results of Figure 5 show that $\gamma$ for $^{15}$NH - $^{15}$NH collisions 
remains $\geq 100$ for all magnetic fields at collision energies above $10^{-3}$ cm$^{-1}$. These results 
indicate that the evaporative cooling of rotationally ground $^{15}$NH molecules in a magnetic trap is likely to be feasible even in the presence of a strong magnetic field. At very low collision energies, $\gamma $ increases dramatically with the field strength, so evaporative cooling to ultracold temperatures would only be possible at B $\leq $ $10^{-3}$ T.
Overall, the present results  support the conclusion of Janssen et al. \cite{JanssenPRA11} that the evaporative cooling of rotationally ground $^{15}$NH molecules in a magnetic trap has a prospect of success.

\section{Summary}

The present work considers the problem of Zeeman relaxation in molecule - molecule collisions in a magnetic field. Accurate calculations of cross sections for elastic scattering and Zeeman relaxation in molecule - molecule collisions are much needed for understanding the prospects of evaporative cooling of molecules in a magnetic trap to ultracold temperatures. Although the theory of molecular scattering in external fields was previously developed \cite{VolpiBohn,Roman04}, the uncoupled space-fixed basis used in all previous studies leads to a prohibitively large number of coupled differential equations. In this work, we show that the cross sections for molecule - molecule collisions can be computed more efficiently using a total angular momentum basis, defined either in the body-fixed frame or in the space-fixed coordinate system. The total angular momentum representation allows for a more physical truncation of the basis set than the fully uncoupled representation, permitting more relevant basis functions to be included in the basis. The fully uncoupled representation \cite{Roman04} became popular for the theoretical description of molecular collisions in external fields because it leads to simple expressions for the Hamiltonian matrix elements that are easy to evaluate. In the present work, we show that compact expressions for the Hamiltonian matrix elements can also be derived in the space-fixed total angular momentum representation. 

We repeated the previous calculations of cross sections for elastic scattering and Zeeman relaxation in $^{15}$NH - $^{15}$NH collisions using the total angular momentum representation leading to a larger basis set. We have obtained converged results demonstrating that the probability of elastic $^{15}$NH - $^{15}$NH scattering remains $>100$ times greater than the probability of Zeeman relaxation at magnetic fields between 10$^{-4}$ and 0.1 T and collision energies  above $10^{-3}$ cm$^{-1}$. 
For collision energies below 10$^{-3}$ cm$^{-1}$, $\gamma $ displays a strong magnetic field dependence and evaporative cooling appears to be feasible only at B $\leq $ $10^{-3}$ T. These results support the conclusions previously reached by Janssen et al. \cite{JanssenPRA11} and remove the uncertainty of the basis set truncation error that constrained the previous results. 

It is well established that low-temperature molecule-molecule scattering properties are extremely sensitive to small variations in the interaction PES. Therefore, while the results presented in Figs. 3 and 5 can be considered converged for a given interaction potential, they are  likely far from being quantitatively accurate. A more careful theoretical analysis is in order to assess the prospects of evaporative cooling of $^{15}$NH in a magnetic trap. Such an analysis can be performed using the methodology outlined in this work and would require extensive computations of thermally averaged ratios of elastic to inelastic collision rates over a wide range of magnetic fields and potential scaling parameters. Because of the large uncertainties in the {\it ab initio} PES, the calculations may quickly become computationally intensive, so it would be best to constrain the $^{15}$NH-$^{15}$NH interaction potential based on future experimental measurements of $^{15}$NH-$^{15}$NH scattering cross sections, as often done for ultracold atoms \cite{Li}.
This will be a challenging task requiring efficient iterative calculations of molecule-molecule scattering cross sections in a magnetic field.

 Molecule-molecule scattering properties determine the possibility of the creation and manipulation of cold molecular gases. Recent theoretical work on dipolar collisions and chemical reactions of ultracold molecules \cite{KRb_theory1,KRb_theory2,KRb_theory3} began to uncover the universal properties of molecule-molecule scattering in the quantum regime. However, not all regimes of molecular collisions are universal, and details of collision dynamics often depend on the intricate interplay between intermolecular and intramolecular
 interactions at short range. Our work complements recent theoretical efforts \cite{KRb_theory1,KRb_theory2,KRb_theory3} by providing an efficient numerical technique for computing the collision properties of polar molecules for arbitrary  collision energies, external field strengths, and interaction potentials. Our proposed methodology can be extended to include the effects of external electric fields, multiple potential energy surfaces and non-adiabatic couplings, and hyperfine interactions, providing a valuable numerical tool for future exploration of molecular collision dynamics in the quantum regime.

\section*{Acknowledgments} 
We thank Liesbeth Janssen, Jeremy Hutson, and Gerrit Groenenboom for discussions and providing the results of their calculations. Y.V.S. and R.V.K. are grateful for the hospitality of the ITAMP visitor program. The computations were perfomed using a computer cluster funded by the Canadian Foundation for Innovation. The allocation of computer time on
Western Canada Research Grid (WestGrid) is also gratefully acknowledged.

\newpage

\newlength{\figwidth}
\setlength{\figwidth}{0.99\columnwidth}

\begin{figure}[ht]
\centering
\resizebox{\figwidth}{!}{\includegraphics[angle=-90]{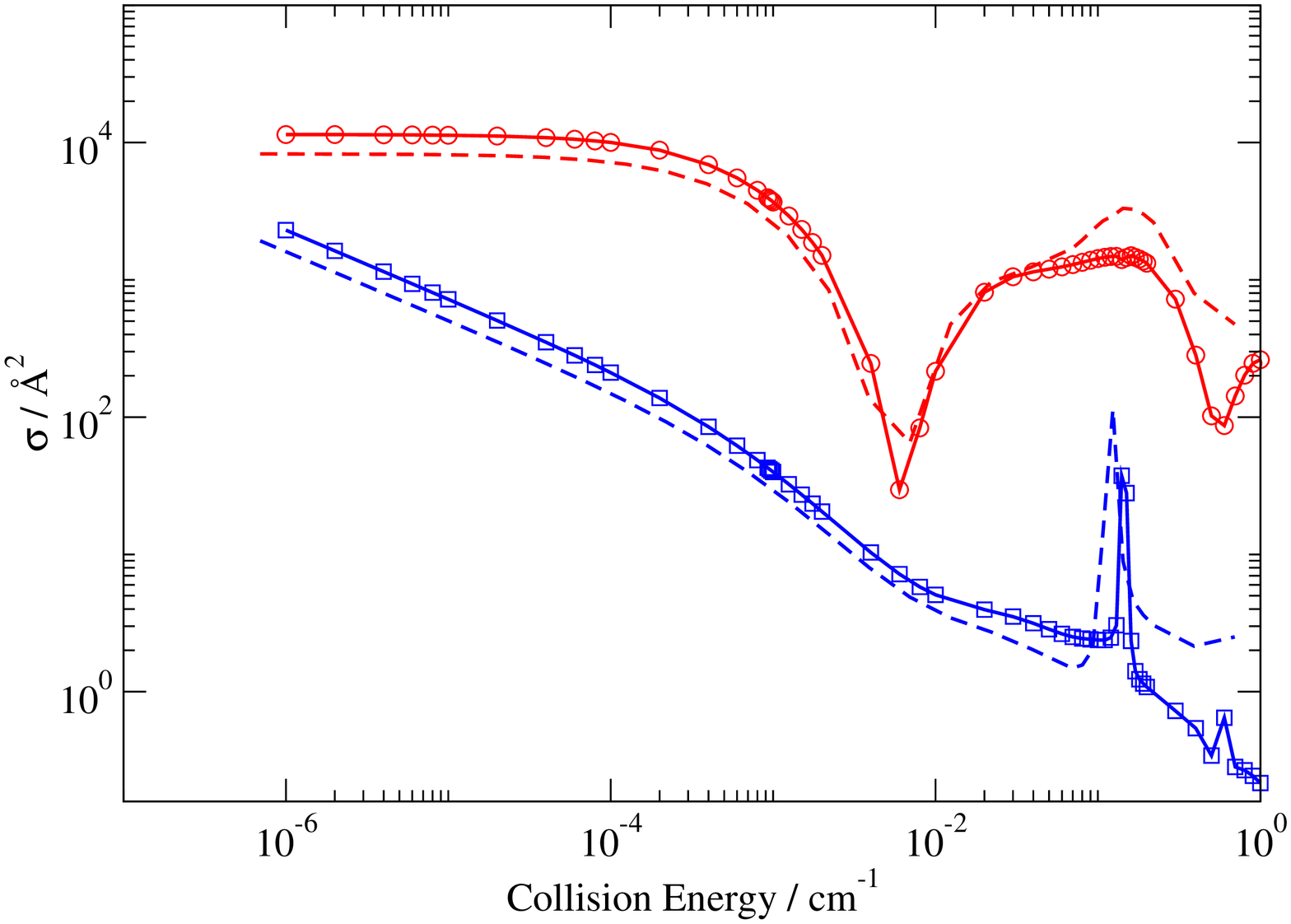}}
\caption{\footnotesize (Color online) Comparison of the total cross sections  for elastic scattering (circles) and spin relaxation (squares) in $^{15}$NH - $^{15}$NH collisions computed using the total angular momentum basis set with $N_{\rm max} = 2$ and $J_{\rm max} = 4$ in this work (solid lines) with the results of Janssen et al. \cite{JanssenPRA11} (dashed lines). The magnetic field is 0.01 T (100G).}
\label{fig:fig2}
\end{figure}

\setlength{\figwidth}{0.99\columnwidth}
\begin{figure}[ht]
\centering
\resizebox{\figwidth}{!}{\includegraphics{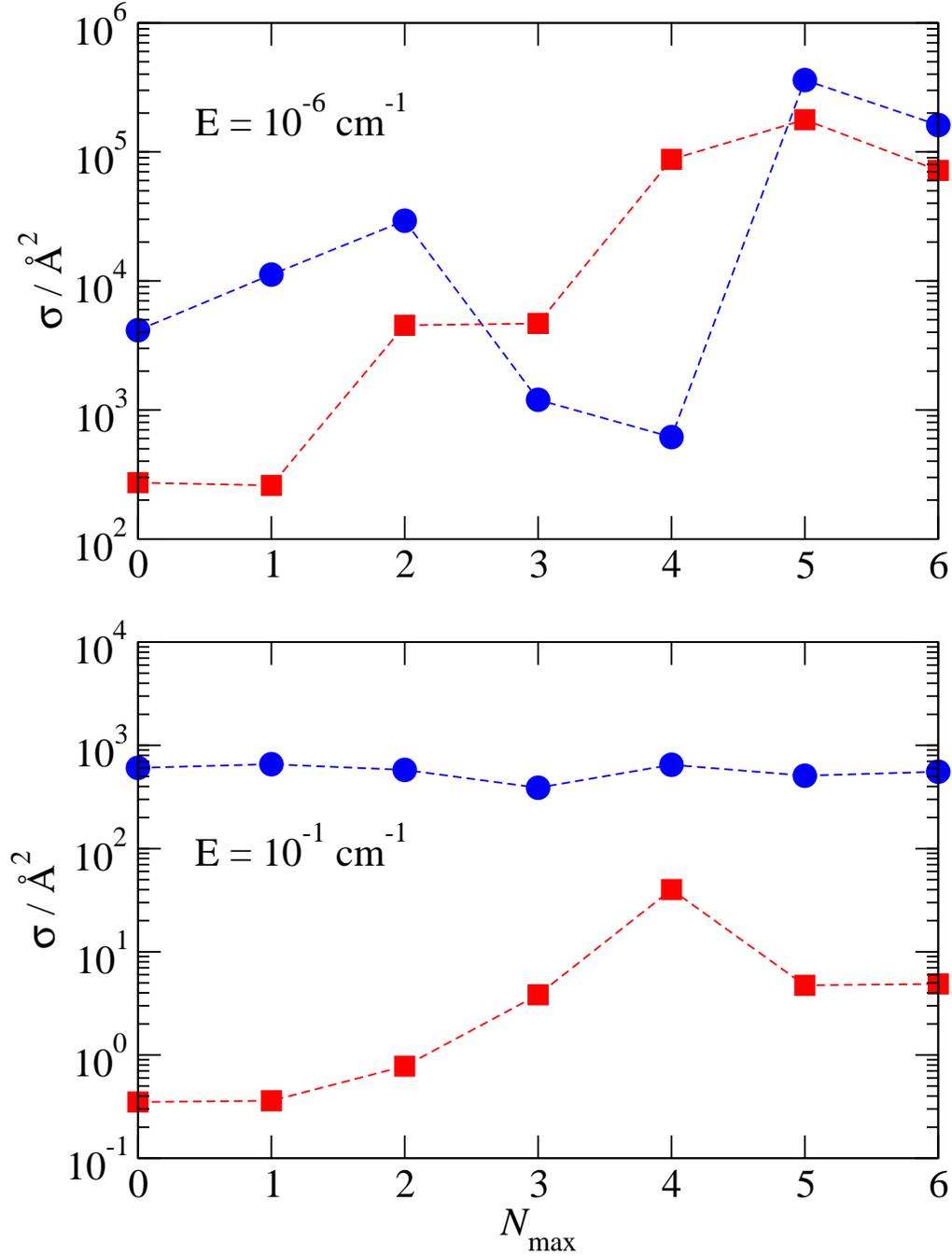}}
\caption{\footnotesize (Color online) Dependence of the cross sections for elastic scattering (circles) and spin relaxation (squares) obtained for $M$ = 2 on the basis set size with the number of rotational states for each molecule restricted by $N_{\rm max}$. The collision energy is 10$^{-6}$ cm$^{-1}$ (upper panel) and 10$^{-1}$ cm$^{-1}$ (lower panel). The magnetic field is 0.1 T (1000 G).}
\label{fig:fig1}
\end{figure}

\begin{figure}[ht]
\centering
\resizebox{\figwidth}{!}{\includegraphics[angle=-90]{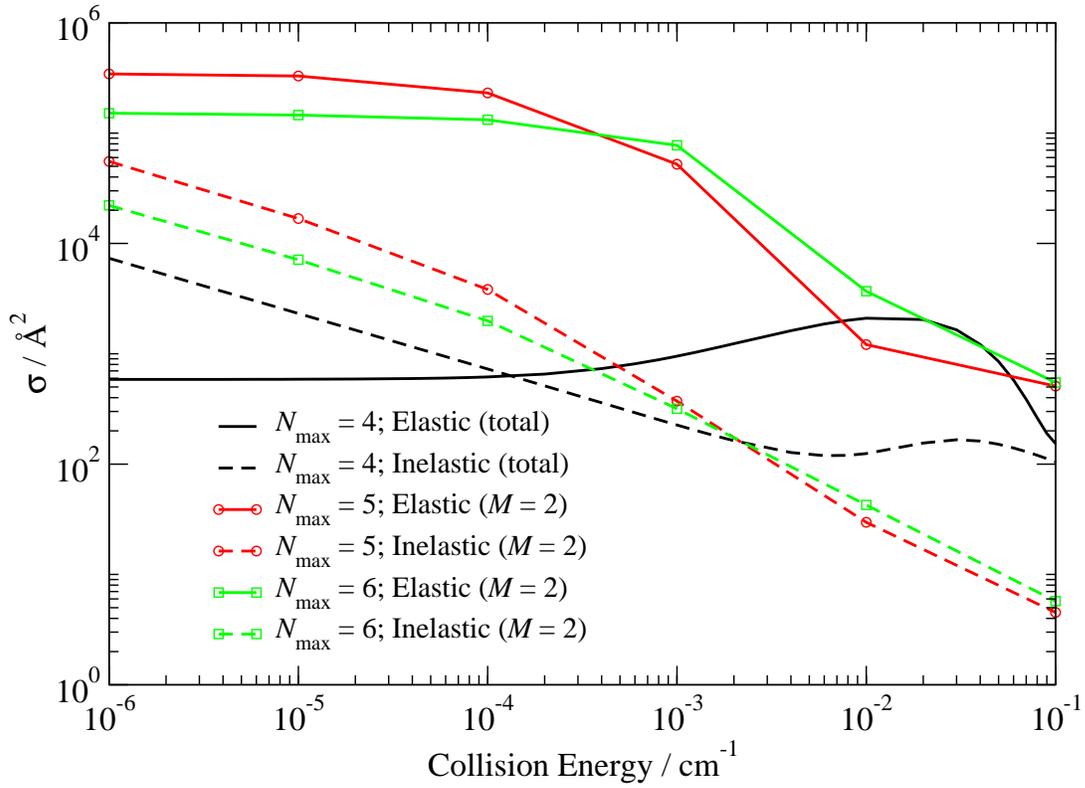}}
\caption{\footnotesize (Color online) The cross sections  for elastic scattering (solid lines) and Zeeman relaxation (dashed lines) as  functions of the collision energy for various basis sets: $N_{\rm max}$ = 4 (no symbols), $N_{\rm max}$ = 5 (circles), and $N_{\rm max}$ = 6 (squares). The magnetic field is 0.01 T (100 G). The calculations for $N_{\rm max}=5$ and 6 were carried out with the fixed value of the total angular momentum projection $M=2$.}
\label{fig:fig3}
\end{figure}

\begin{figure}[ht]
\centering
\resizebox{\figwidth}{!}{\includegraphics[angle=-90]{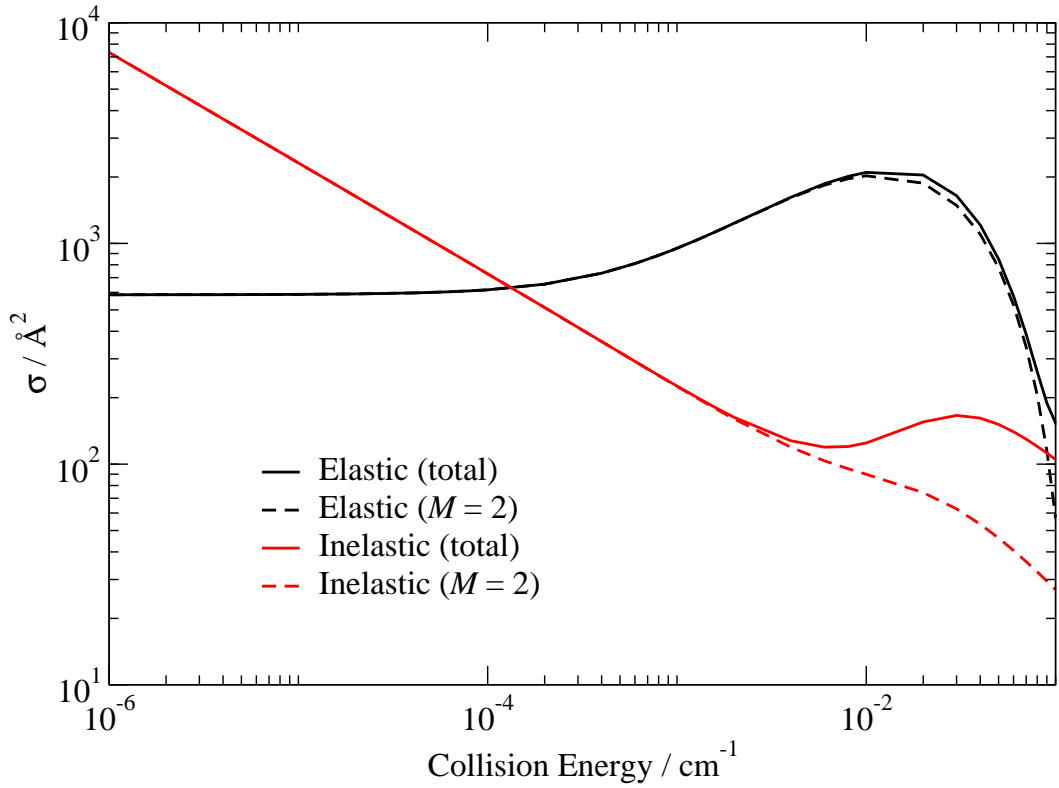}}
\caption{\footnotesize (Color online) The cross sections for ealstic scattering and Zeeman relaxation in $^{15}$NH - $^{15}$NH collisions computed after summation over all possible values of $M$  (solid lines) and with the fixed value $M = 2$ (dashed lines). The basis set includes the rotational states up to $N_{\rm max}$ = 4 and total angular momentum states up to $J_{\rm max}$ = 4 for each molecule. The magnetic field is 0.01 T (100 G).}
\label{fig:fig4}
\end{figure}

\begin{figure}[ht]
\centering
\resizebox{\figwidth}{!}{\includegraphics[angle=-90]{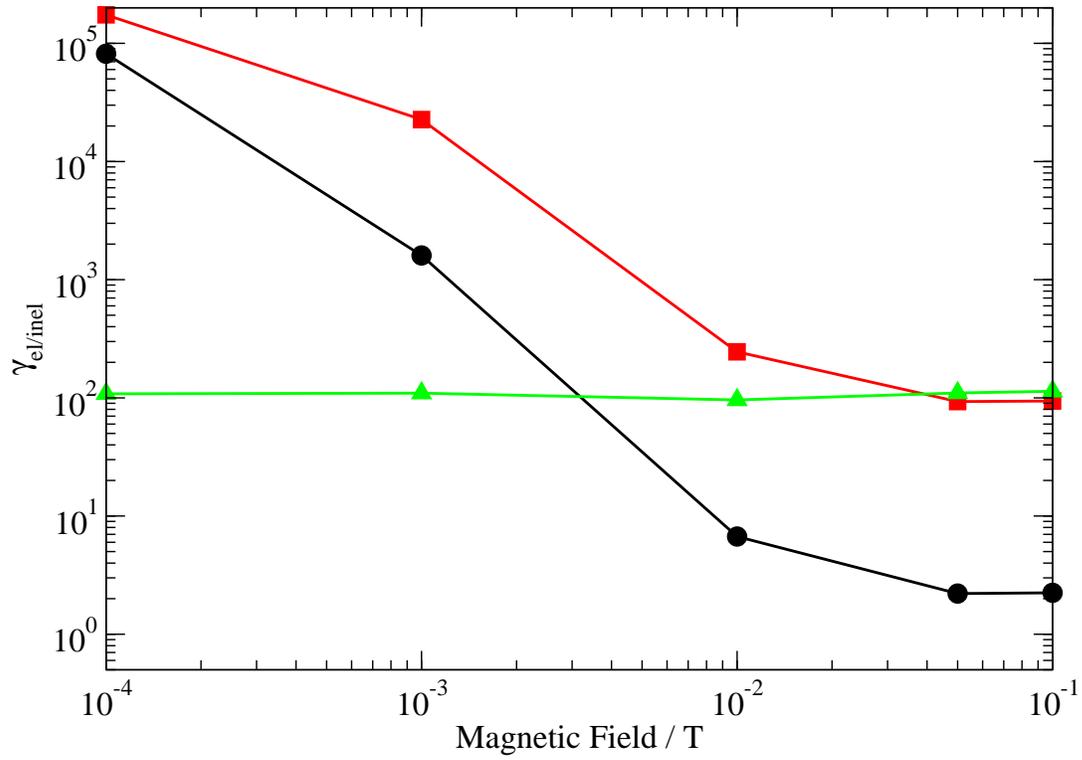}}
\caption{\footnotesize (Color online) The ratios of the converged ($N_{\rm max}=6$) cross sections for elastic scattering and Zeeman relaxation in $^{15}$NH-$^{15}$NH collisions  as functions of the magnetic field at different collision energies:  10$^{-6}$ cm$^{-1}$ (circles), 10$^{-3}$ cm$^{-1}$ (squares) and 0.1 cm$^{-1}$ (triangles).}
\label{fig:fig5}
\end{figure}

\end{document}